\documentclass[12pt,preprint]{aastex}

\shorttitle{Chandra discovery of a binary AGN in NGC\,6240}
\shortauthors{St. Komossa et al.}

\begin{document}

\title{ {\small{ApJ Letters, in press (scheduled for January 1, 2003)}} \newline
            \mbox{    }   \newline
Discovery of a binary AGN in the 
 ultraluminous infrared galaxy NGC\,6240 using {\itshape Chandra} }

\author{St. Komossa$^{1}$, V. Burwitz$^{1}$, G. Hasinger$^{1}$, 
            P. Predehl$^{1}$, J.S. Kaastra$^{2}$, Y. Ikebe$^{3}$}
\affil{$^{1}$Max-Planck-Institut f\"ur extraterrestrische Physik,
Giessenbachstrasse 1, 85748 Garching, Germany, skomossa@xray.mpe.mpg.de; 
$^{2}$SRON National Institute for Space Research, Sorbonnelaan 2, 3584 Utrecht, The Netherlands;
$^{3}$Joint Center for Astrophysics,
Physics Department,
University of Maryland, Baltimore County,
1000 Hilltop Circle,
Baltimore, Maryland 21250, USA }

\begin{abstract}
Ultraluminous infrared galaxies (ULIRGs) are outstanding
due to their huge luminosity output in the infrared,
which is predominantly powered by super starbursts and/or hidden active galactic
nuclei (AGN).
NGC\,6240 is one of the nearest ULIRGs and is considered a key
representative of its class. 
Here, we report the first high-resolution imaging spectroscopy 
of NGC\,6240 in X-rays.  
The observation, performed with the ACIS-S detector aboard
the {\sl Chandra} X-ray observatory, led to the discovery of 
two hard nuclei, coincident with the optical-IR nuclei of NGC\,6240. 
The AGN character of {\em both} nuclei is revealed by
the detection of absorbed hard, luminous X-ray emission and
two strong neutral Fe\,K$\alpha$ lines.  
In addition, extended X-ray emission components are present,
changing their rich structure 
in dependence of energy. 
The close correlation of the extended emission with the optical 
H$\alpha$ emission of NGC\,6240, 
in combination with the softness of its spectrum, clearly indicates
its relation to starburst-driven superwind activity. 

\end{abstract}


\section{Introduction}

Ultraluminous infrared galaxies (ULIRGs) are characterized by  
their huge luminosity output in the infrared,
which is predominantly powered by super-starbursts and/or hidden AGN
(e.g., Genzel et al. 1998, Sanders 1999).
Many distant {\sl SCUBA} sources, massive and dusty galaxies,
are believed to be ULIRG equivalents at high redshift
(e.g., Barger et al. 1998, Lawrence 2001).
Local ULIRGs, of which NGC 6240 is regarded a key representative, 
are therefore important
laboratories to study the physics of superwinds
driven by the nuclear starbursts,
the processes of IGM enrichment,
to search for  the presence of
hidden AGN,
and to study the physics of galaxy formation
(many ULIRGs are interacting galaxies, believed to be on
the verge of forming elliptical galaxies).

NGC\,6240 is one of the nearest members
of the class of ULIRGs, and has been intensely studied
at virtually all wavelengths.
The galaxy shows conspicuous loops and tails,
and two optical nuclei separated by $\sim$1.8$^{\prime\prime}$
(Fried \& Schulz 1983) which are also detected in the IR and radio band.  
Recent observations suggest that NGC\,6240 is in a relatively
early merger state (Tacconi et al. 1999, Tecza et al. 2000, Gerssen et al. 2001). 
It is expected to finally form an elliptical galaxy. 
Ground-based optical spectroscopy of NGC\,6240 shows LINER-like 
emission-line ratios. 

Given the importance of the question whether an AGN is
present, and whether it is the ultimate power-source of the ULIRG NGC\,6240,
efforts to search for such an AGN were undertaken at all wavelengths
(see Sect. 1 of Komossa et al. 1998 for a review). 
Recent observations include the detection of flaring water-vapor
maser emission from NGC\,6240 (Nakai, Sato \& Yamauchi, 2002), 
the presence of a high-excitation [OIV] infrared emission line
(Genzel et al. 1998),
and the compactness of the radio cores 
(Beswick et al. 2001). 
In X-rays, NGC\,6240 exhibits exceptionally
luminous extended
emission (Schulz et al. 1998,
Komossa et al. 1998, Kolaczyk \& Dixon 2000) likely powered by superwinds, 
and an iron-line superposed on 
hard X-ray emission (Mitsuda et al. 1995, Schulz et al. 1998,
Iwasawa \& Comastri 1998) extending beyond 10 keV (Vignati et al. 1999,
Ikebe et al. 2000), interpreted as arising from an absorbed active nucleus. 
It is the hard X-ray observations that 
give the best evidence for the presence 
of an AGN in
NGC\,6240. 

It remained basically an open question, which of the two nuclei of NGC\,6240 is 
the active one (or whether even both of them are);
{\sl HST} observations
(Rafanelli et al. 1997) suggested that the southern nucleus harbors
an AGN or a heavily obscured LINER.

The questions regarding the onset of starburst and AGN activity and their
evolution in mergers are of fundamental importance for our understanding
of AGN/black hole formation and evolution in general. 
Given the complex nature of the X-ray emission of NGC\,6240 with
contributions from many components suggested from previous
X-ray observations (Schulz et al. 1998, Komossa et al. 1998,
Iwasawa \& Comastri 1998, Netzer et al. 1998, Vignati et al. 1999,
Ikebe et al. 2000), 
spatially resolved spectroscopy is crucial in order for us to disentangle
all contributing components, determine their nature, and derive their
physical properties. 
In this {\sl Letter}, we report the {\sl Chandra} 
discovery that {\em both} nuclei of NGC\,6240 are active,
and we show first results on the remarkably structure-rich extended X-ray 
emission.

Luminosities were calculated using a Hubble constant of 
$H_0 = 50$ km s$^{-1}$ Mpc$^{-1}$. At a distance of 
NGC\,6240 of 144 Mpc, 1$^{\prime\prime}$ corresponds to
700 pc in the galaxy.

\section{X-ray observations and data analysis}

NGC\,6240 was observed as part of the guaranteed time
observer program with the Advanced CCD Imaging
Spectrometer (ACIS-S) instrument 
onboard the {\sl Chandra} X-ray observatory. 
The observation was carried out on July 29, 2001
with an effective exposure time of 37 ks. 
The pointing of the telescope was such 
that the X-ray photons of the target source were
registered on the back-illuminated S3 chip of ACIS.  
The data analysis was prepared using the CIAO2.2
software tool. 

Spectra were grouped 
to have at least 25 photons per bin. 
Since few photons were detected below 0.5 keV, and since the
instrument response is still somewhat uncertain at
low energies, we restricted our modeling to photon
energies above 0.5 keV. 

The ACIS-S background is very low. 
It was determined in a circular, source-free region  
north of the central X-ray emission, and was subtracted. 
In order to carry out the spectral analysis of each
nucleus separately, source photons
from the northern and southern nucleus were extracted in circular regions of 
radii 0.8$^{\prime\prime}$ and 0.9$^{\prime\prime}$ respectively,
centered at the X-ray positions
of the nuclei. 
Further details on photon extraction areas for
other regions of NGC\,6240 are given below.  

\section{X-ray morphology}

The {\sl Chandra} image of NGC\,6240 reveals a wealth of structure,
changing in dependence of energy (Fig. 1). 
A large part of the X-ray emission of NGC\,6240 is extended, confirming previous
results from {\sl ROSAT} (Komossa et al. 1998, Schulz et al. 1998) and a short
{\sl Chandra} HRC-I observation
(Lira et al. 2002). The previous observations did not provide
any energy information, though.

Several X-ray `loops' and knots are visible that correlate
well with the H$\alpha$ emission (Keel 1990, Gerssen et al. 2001)
of NGC\,6240. 
Apart from the complex central emission, only one X-ray source
is detected within the optical confines of NGC\,6240. 
Its spectrum is rather hard. The source 
could be a background AGN. 
In particular, the following features stand out in energy images
of NGC\,6240:

(0.1-3.0)\,keV: 
Below 0.5 keV, barely any X-ray photons are detected.
This can be traced back to the high absorption toward
NGC\,6240. Above 0.5 keV,
extended, loop-like emission and several
knots appear, prominent below 2.5 keV. 
Of special interest is the `blob' south-west of the southern nucleus,
which appears above 1 keV (Fig. 1, Fig. 2). It is the strongest 
feature between 1.0 and 1.5 keV and is spatially extended. 
Inspecting {\sl HST} images (Gerssen et al. 2001), we find it correlates with a region
of increased H$\alpha$ emission, and may correspond
to a more recent superwind outflow than the more widely extended structures. 
Above 1.5 keV, X-ray emission from
the direction of the northern nucleus of NGC\,6240 starts to emerge.  

(5.0-8.0)\,keV: The hard X-ray image is dominated by emission
from two compact sources (Fig. 2, 3), spatially coincident within the errors
with the IR position of the two nuclei of NGC\,6240.
With a distance of 1.4 arcseconds,
the X-ray nuclei are slightly closer together than the optical nuclei (Fig. 2), 
consistent with
the previous interpretation that extinction causes a wavelength-dependent
shift in the flux centroids of the nuclei (Schulz et al. 1993).  
Both nuclei show emission from the neutral iron line.
The southern nucleus is more  prominent in this line than the northern nucleus.
Running a standard source detection routine, both nuclei are detected as
sources in the energy interval 5-8\,keV with signal/noise ratios
of 9.9 (southern nucleus) and 3.5.  
 
We do not detect any X-ray emission from the location of the 
steep off-nuclear velocity gradient (Bland-Hawthorn, Wilson \& Tully 1991,
Gerssen et al. 2001), 
which was recently suggested to be possibly due to
a kinematic gradient in a starburst wind.  
No correlation of X-ray emission with the radio-arm west of the nuclei,
described by Colbert et al. (1994) and suggested to be
linked to superwind activity, was found.
Finally, no X-ray point source was detected from the direction
of the supernova SN\,2000bg.

\section{X-ray spectroscopy}

\subsection{Extended emission}

Representative of the extended emission, we report here results
of spectral fits to the north-eastern-loop extended emission (Fig. 1). 
Source photons were extracted from an elliptical region
centered at RA=16:52:59.3, DEC=02:24:08.1 (ellipse radii: 8.9$^{\prime\prime}$
and 10.6$^{\prime\prime}$, angle=45$^{\rm o}$).  
We find that the X-ray spectrum is well described
by a MEKAL model (Mewe et al. 1995) with $kT=0.81\pm{0.05}$ keV and absorption with
column density $N_{\rm H} = (3.1{\pm{0.4}})\,10^{21}$ cm$^{-2}$
($\chi^{2}_{\rm red}=1.2$),
which exceeds the Galactic value in the direction 
of NGC\,6240, $N_{\rm Gal} = 5.49\,10^{20}$ cm$^{-2}$,
and is consistent with optical estimates of the extinction toward
the extended optical emission of $A_{\rm v} \approx$ 1 mag (Thronson et al. 1990).   
Metal abundances approach 0.1$\times$solar. 

Secondly, the `blob' south-west of the  southern nucleus was
fit after extracting its photons from a circular
region centered at RA=16:52:58.8, DEC=02:24:1.5. 
Its spectrum is significantly harder and more heavily absorbed. 
Spectral fits consisting of a two-component MEKAL model were
applied. Among these two components, one was assumed to be the
extension of the widely extended soft X-ray emission of NGC\,6240
part of which is projected
onto the central region of NGC\,6240. Explicitly, its parameters
were all fixed to those determined for the north-eastern-loop 
extended X-ray emission (including abundances and the amount of absorption),
apart from the normalization
that was treated as free parameter. 
The second, hotter, MEKAL component was added to describe 
the X-ray emission from the south-western blob itself. 
The spectrum of the south-western blob is then well described 
by a MEKAL model with $kT=2.8\pm{0.8}$ keV and $N_{\rm H} = (7{\pm{4}})\,10^{21}$ cm$^{-2}$
($\chi^{2}_{\rm red}=0.95$).
Abundances are not well constrained, but are close to 0.8$\times$solar and were
thus fixed to that value. 

\subsection{Nuclear emission}

Two circular regions centered at the southern and northern nucleus
were selected for spectral analysis. 
The residual soft starburst emission projected onto
the nuclei (Fig. 1) was fit by a MEKAL model, 
its normalization and temperature 
left free to vary. The derived temperature ($kT=0.9\pm{0.2}$\,keV)
is consistent within the errors
with that of the north-eastern loop emission.
The extinction toward the nuclei
estimated from IR observations, $A_{\rm v}^{\rm S}$=5.8 mag and
$A_{\rm v}^{\rm N}$=1.6 mag (Tecza et al. 2000), corresponds
to absorption column densities of $N_{\rm H}^{\rm S}$ = 10$^{22}$ cm$^{-2}$ 
and $N_{\rm H}^{\rm N}$ = 2.8\,10$^{21}$ cm$^{-2}$,
and is expected to be the minimum amount of absorption along
the line of sight towards the nuclei. MERLIN radio observations of neutral
hydrogen toward the radio nuclei of NGC\,6240 indicate column densities
of order $N_{\rm H}^{\rm S,N} \approx (1.5-2)\,10^{22}$ cm$^{-2}$ (Beswick et al. 2001).  

Both nuclei show very hard X-ray spectra extending out to $\sim$ 8-9\,keV.
If the spectra are fit by a single powerlaw, strong residuals
around the location of the neutral 6.4\,keV iron K$\alpha$ line are
visible, and further weaker ones around $\sim$6.95\,keV. The energy of the second line
is close to H-like iron and K$\beta$ of neutral iron.  
The southern nucleus is brighter in X-rays than the northern 
nucleus. Its spectrum is well fit by a very flat powerlaw
of photon index (defined as in $F \propto E^{+\Gamma}$) 
$\Gamma_{\rm x}=-0.2\pm{0.3}$,
and two narrow iron lines at energies corresponding to neutral and H-like iron
(Fig. 4). The emission is dominated by the neutral iron line that 
is located at a (redshift-corrected) energy $E$=6.42$\pm{0.03}$ keV, with 
 1.5\,10$^{-5}$  ph\,cm$^{-2}$\,s$^{-1}$. 
The amount of excess absorption is not well constrained. It is of order
$N_{\rm H}^{\rm S} \simeq 10^{22}$ cm$^{-2}$ and is uncertain by at least a factor 2. 
The spectrum of the northern nucleus is less hard and less absorbed.
It is well described by a powerlaw of photon index $\Gamma_{\rm x}=-0.9\pm{0.2}$,
absorbed with $N_{\rm H}^{\rm N}=(0.6\pm{0.2})\,10^{22}$ cm$^{-2}$ in addition to
the Galactic $N_{\rm Gal}$, and a narrow FeK$\alpha$ line
of neutral iron (Fig. 4) that is about a factor 3 weaker than in the
southern nucleus.
The absorption corrected fluxes in the (0.2-10)\,keV band 
are $f_{\rm x, S} = 7.6\,10^{-13}$ erg\,cm$^{-2}$\,s$^{-1}$
and $f_{\rm x, N} = 2.7\,10^{-13}$ erg\,cm$^{-2}$\,s$^{-1}$
for southern and northern nucleus,
respectively.

\section{Discussion}

\subsection{Starburst-related emission} 

Could the widely extended X-ray emission of NGC\,6240
still be powered by photoionization 
of the active nuclei ?  {\sl Chandra} recently found some cases in which 
photoionization seems to dominate the ionization of
circum-nuclear matter out to large distances
($\sim$ 2\,kpc in case of NGC\,1068; Young, Wilson \& Shopbell 2001). 
However, the correlation of X-ray and H$\alpha$ emission in combination with the 
{\em soft}, thermal-plasma-like spectrum of the extended emission of NGC\,6240
strongly favors starburst-driven superwinds (Heckman, Armus \& Miley 1987)
as the origin for the bulk of the extended X-ray emission
of NGC\,6240. 
According to superwind models of Schulz et al. (1998),
a mechanical input power of $L_{\rm mech} = 3\,10^{43}$ erg/s
(derived from a SN-rate of 3/year) can, within 3\,$10^7$ yrs, 
drive a single shell to an extent of $R \sim$ 10\,kpc 
within a medium of 0.1 cm$^{-3}$ particle density.
This model reproduces the observed X-ray luminosity
of the extended emission of $\sim$10$^{42}$ erg/s.  
A distance of 10 kpc corresponds to the extent of the five-finger structure
seen in H$\alpha$, coincident with the brightest extended X-ray emission,
including a correction for a disk inclined by up to 40$^{\rm o}$ from
edge-on.

The ionized iron line observed in addition
to the neutral iron line may originate in `Cas-A-type'
supernova remnants (e.g., Willingale et al. 2002), 
or in a near-nuclear `warm scatterer' that is ionized
by the AGN (Komossa et al 1998).

The extended soft X-ray component and its link
with non-X-ray morphological structures will be further discussed 
in a follow-up
paper (Komossa et al., in prep.). In the following, we concentrate on the
core region of NGC\,6240.

\subsection{Emission from the binary AGN}

Using {\sl Chandra} ACIS-S, we found that {\em both} nuclei
of NGC\,6240 are emitters of luminous hard X-ray emission,
on which a strong neutral iron K$\alpha$ line is superposed. 
These properties
identify both nuclei of NGC\,6240 as active. 
In particular, a strong, neutral Fe K$\alpha$ line is not produced
in a starburst-superwind, but originates from fluorescence in
cold material illuminated by a hard continuum spectrum.  
For the first time, we can disentangle the contribution to
the hard X-ray luminosity from the southern and northern nucleus.
The observed (0.1-10)\,keV, absorption-corrected, fluxes 
convert to {\em observed} X-ray luminosities of 
the southern and northern nucleus, $L_{\rm x, S}=1.9\,10^{42}$ erg/s,
$L_{\rm x, N}=0.7\,10^{42}$ erg/s,
respectively.
 
The {\em intrinsic} luminosities are significantly larger,
because 
it has been repeatedly argued that the emission
we are seeing in X-rays below 10 keV is scattered emission
from the AGN (e.g., Mitsuda et al. 1995, Schulz et al. 1998,
Komossa et al. 1998).
The intrinsic AGN emission of NGC\,6240 only shows
up above $\sim$9-10 keV (Vignati et al. 1999, Ikebe et al. 2000). 
The strength of the neutral Fe K$\alpha$ line in both nuclei
suggests a scattering geometry for {\em both} AGN.
This is also indicated by the flatness of the hard powerlaw spectra (Sect. 4.2).  

We find marginal evidence that part of the
hard emission is extended by few arcsesonds.
3$^{\prime\prime}$ correspond to $\sim$2 kpc in NGC\,6240.
It is interesting to note that this scale is
similar to the extent of the hard X-ray emission and iron line 
of  NGC\,1068 detected with {\sl Chandra} (Young, Wilson \& Shopbell 2001). 
Dense cold gas that could provide the scattering agent is
abundant in the center region of NGC\,6240. 
Deeper {\sl Chandra} observations of NGC\,6240, still
employing the superb imaging spectroscopy capabilities
of the ACIS-S detector, will be indispensable 
for studying the complex central region of NGC\,6240 in greater
detail; for instance to obtain a high-quality 
image in the Fe\,K$\alpha$ line. 

The presence of supermassive binary black holes, as in NGC\,6240, are
of importance for our understanding of the formation and evolution
of AGN and the formation of elliptical galaxies via mergers.

Ultimately, the binary AGN of NGC\,6240 will
coalesce to form one nucleus. 
The final merging of the supermassive black holes is expected to produce
a strong gravitational wave signal. In fact, such events
are expected to generate the clearest  
signals detectable with 
the gravitational wave detector LISA that will be placed in Earth orbit in the near future 
(e.g., Danzmann 1996).

\acknowledgments

It is a pleasure to thank Reinhard Genzel for his comments and encouragement, Elisa Costantini
for discussions about the {\sl Chandra} analysis of extended sources, and an anonymous referee for
his/her many useful comments and suggestions.

\clearpage


\begin{figure}
\begin{center}
\begin{minipage}[t]{14cm}
\includegraphics[width=7.0cm, angle=0]{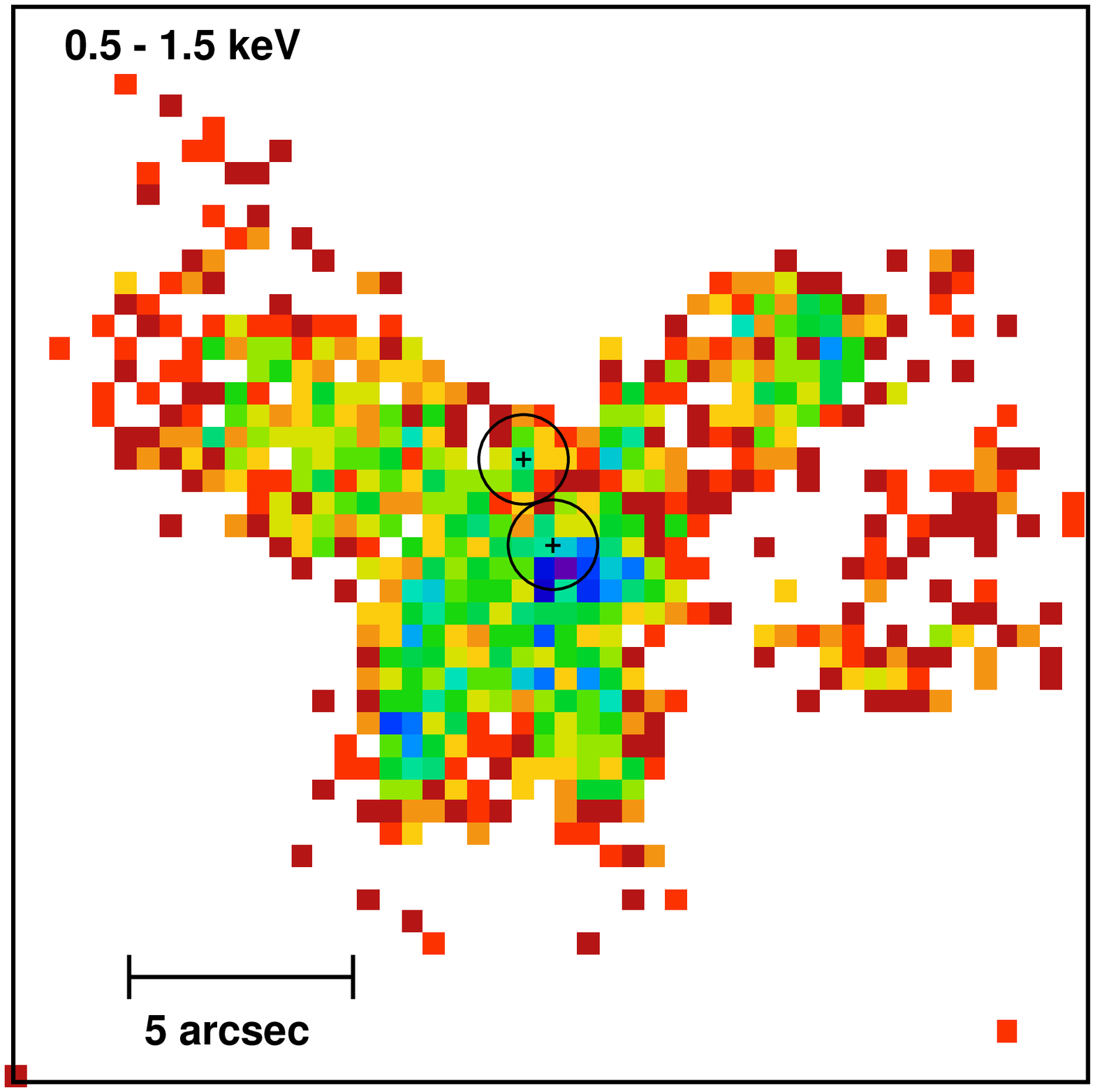}
\includegraphics[width=7.0cm, angle=0]{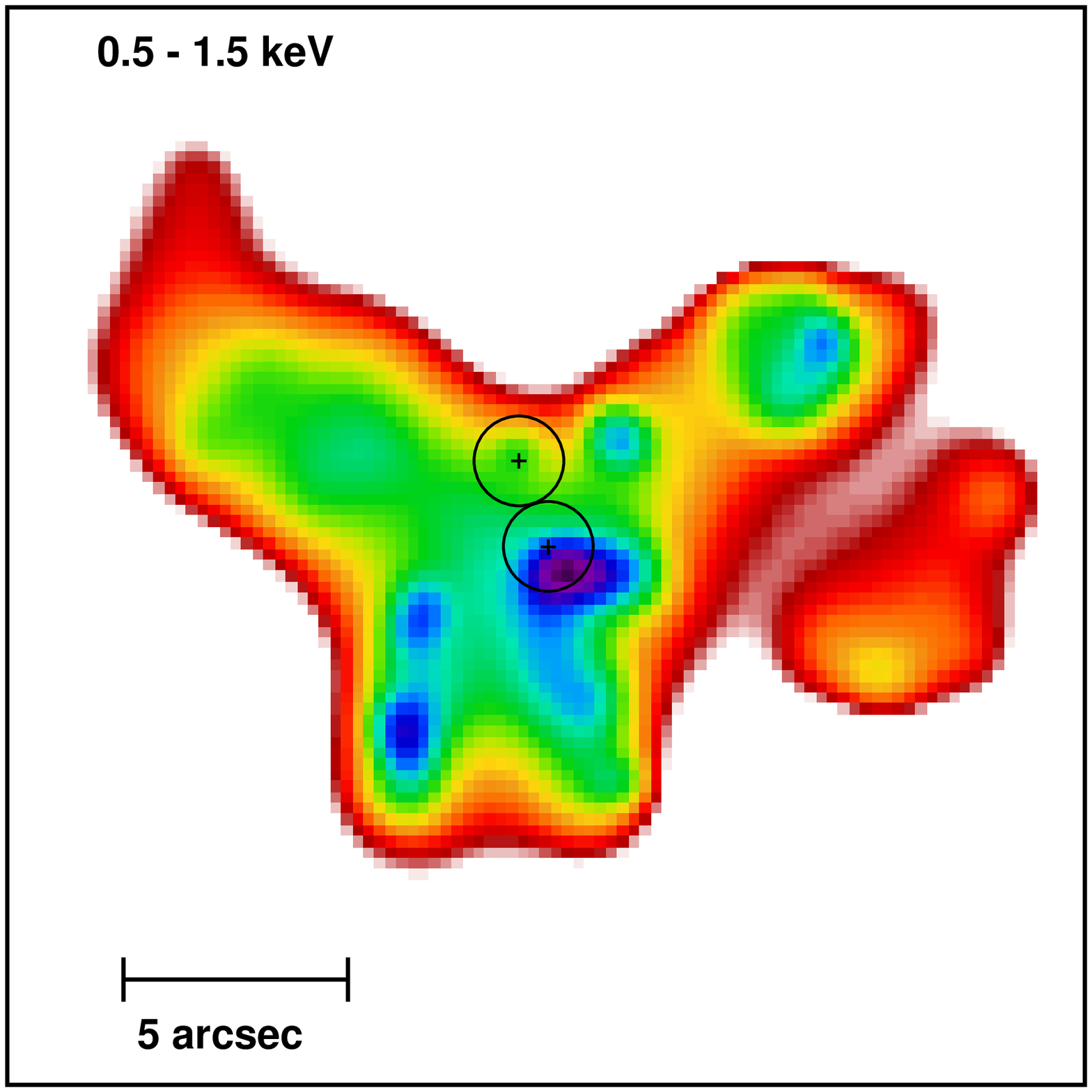}
\includegraphics[width=7.0cm, angle=0]{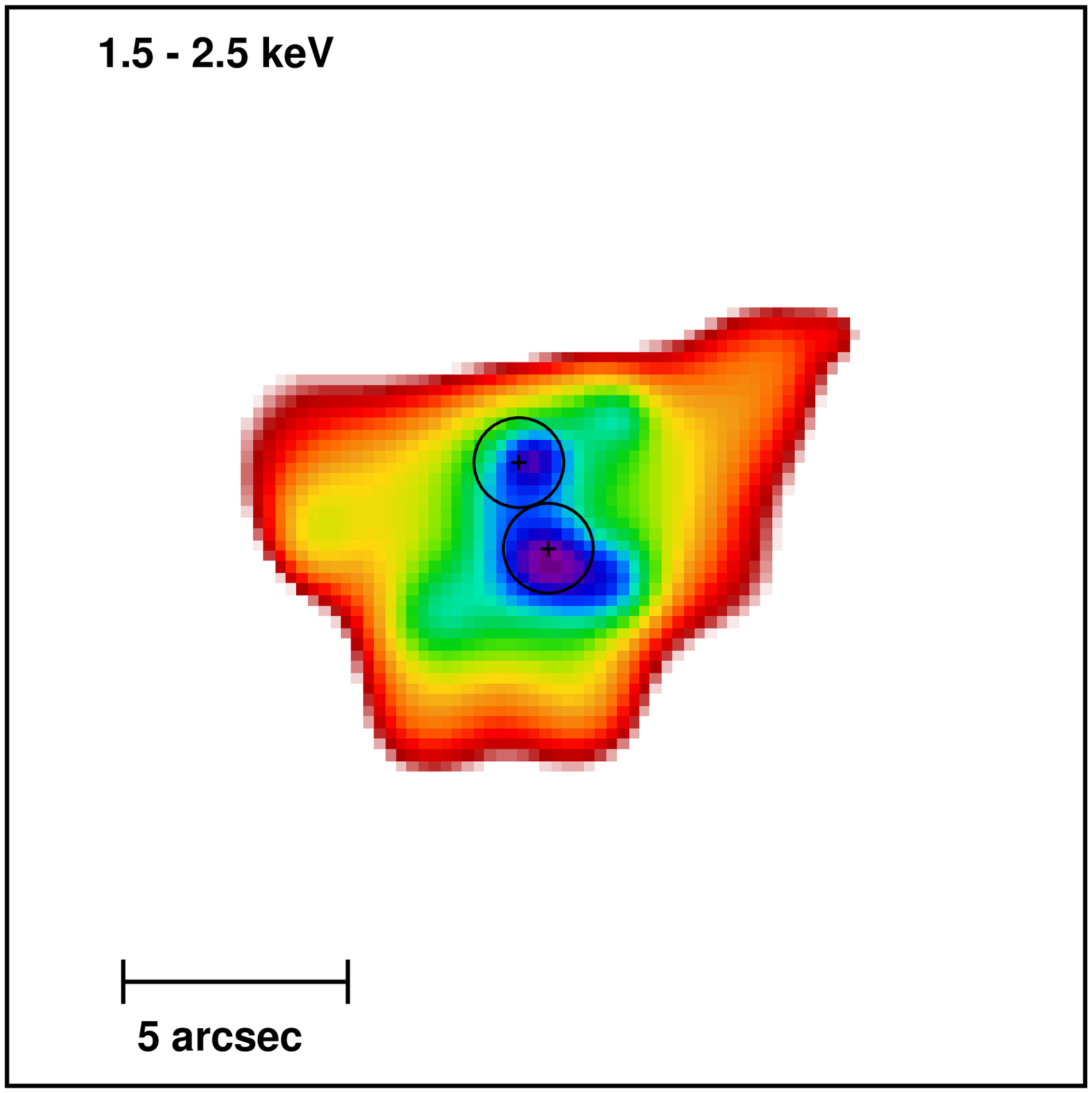}
\includegraphics[width=7.0cm, angle=0]{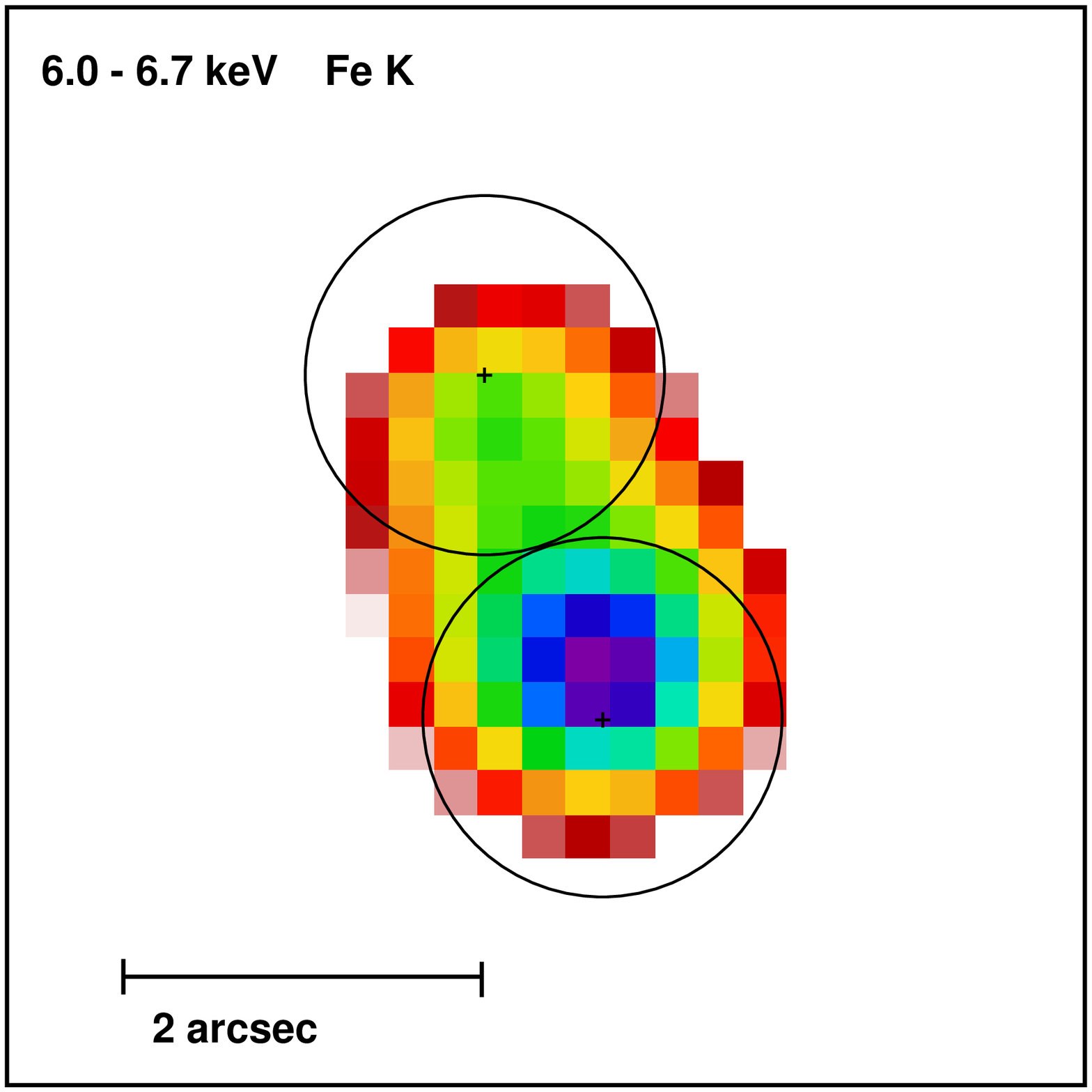}
\end{minipage}
\end{center}
\caption{ X-ray images of NGC\,6240 in selected energy intervals. From top to bottom:
The first image shows the raw ACIS-S image in the energy range 0.5-1.5 keV.
The circles mark the optical positions of the two nuclei of
NGC\,6240. 
Further images are shown in selected energy intervals,
as marked in the figures.    
The images were adaptively smoothed using the FFT algorithm in the CSMOOTH
routine available in CIAO 2.2.  }
\end{figure}

\clearpage 

\begin{figure}
\plotone{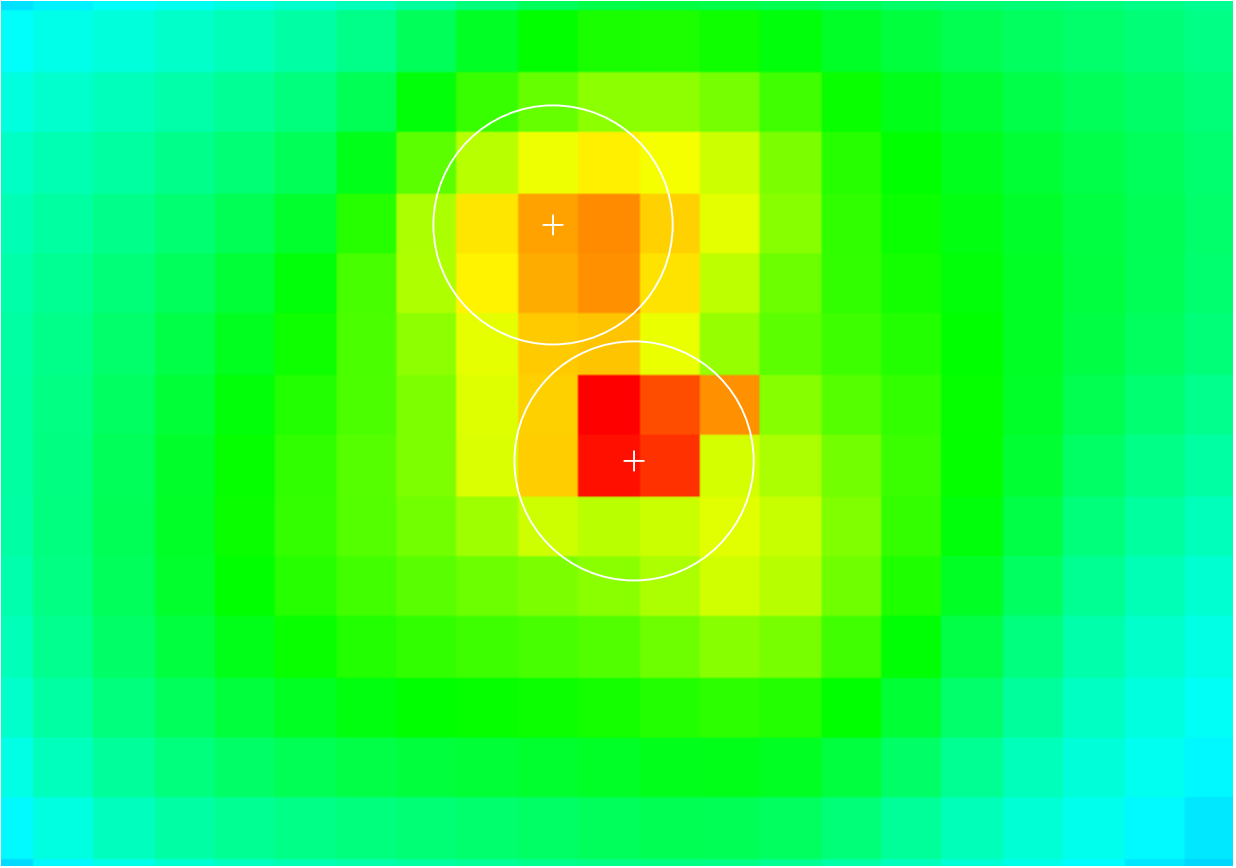} 
\caption{ X-ray image of NGC\,6240 in the energy range 2.5--8 keV,
zoomed on the two nulcei. The circles mark the optical positions 
of the nuclei. 
}
\end{figure}

\clearpage 

\begin{figure}
\begin{center}
\begin{minipage}[t]{16cm}
\includegraphics[width=8.0cm, angle=90]{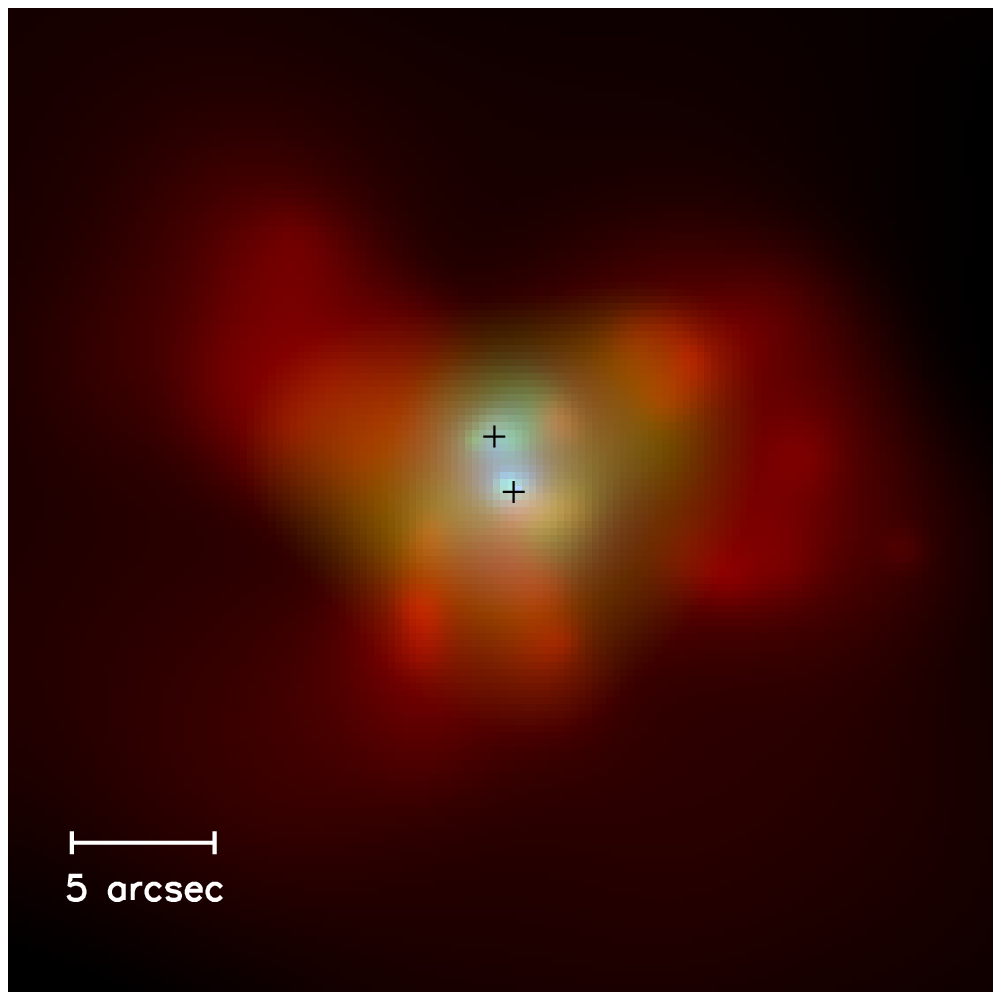}
\includegraphics[width=8.0cm, angle=90]{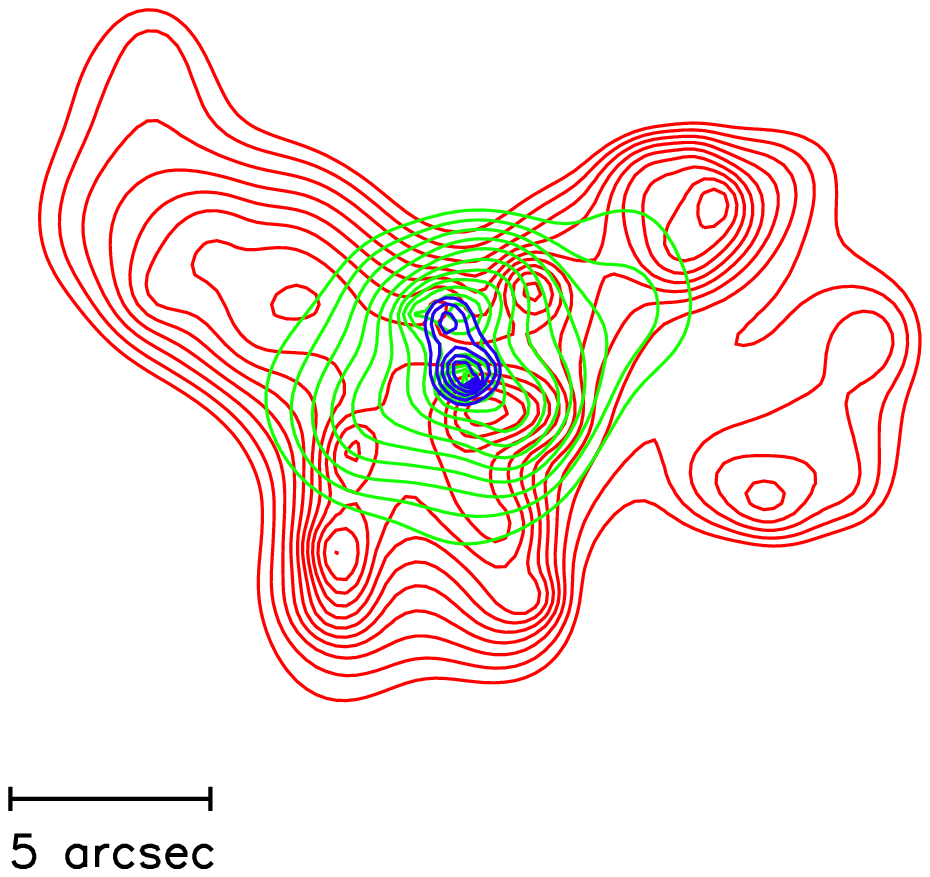}
\end{minipage}
\end{center}
\caption{Multi-colour image of NGC\,6240. Red=soft (0.5-1.5 keV), green = medium (1.5-5 keV)
and blue = hard (5-8 keV) X-ray band.  The right image shows  
contour plots, using the same colour coding.  
 }

\end{figure}

\clearpage

\begin{figure}
\plotone{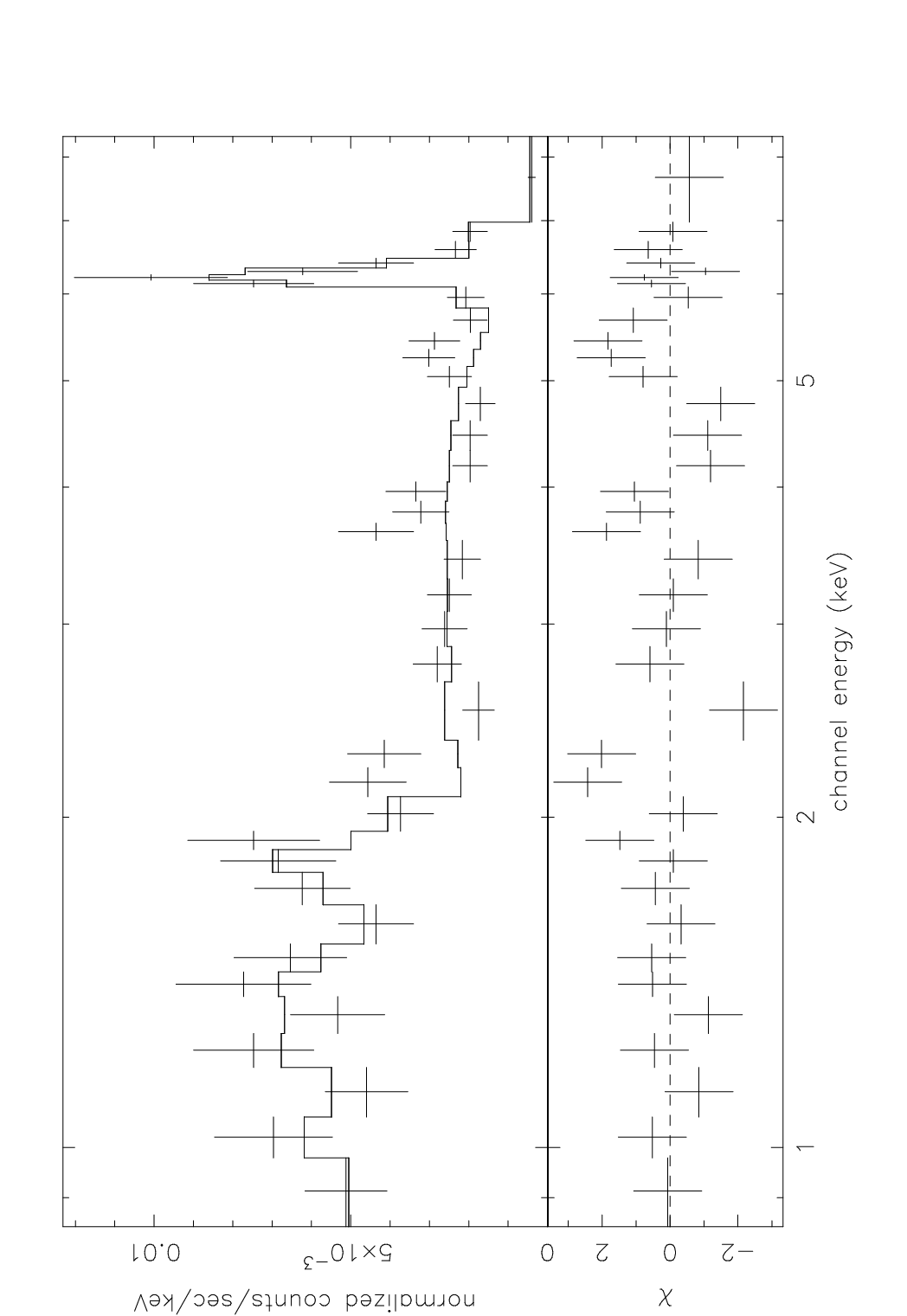}
\caption{ X-ray spectrum of  southern
nucleus of NGC\,6240. A model consisting of thermal emission plus an absorbed powerlaw
and two Gaussian emission lines 
was fit to the data.  The lower panels show the residuals. }
\end{figure}

\clearpage

\begin{figure}
\plotone{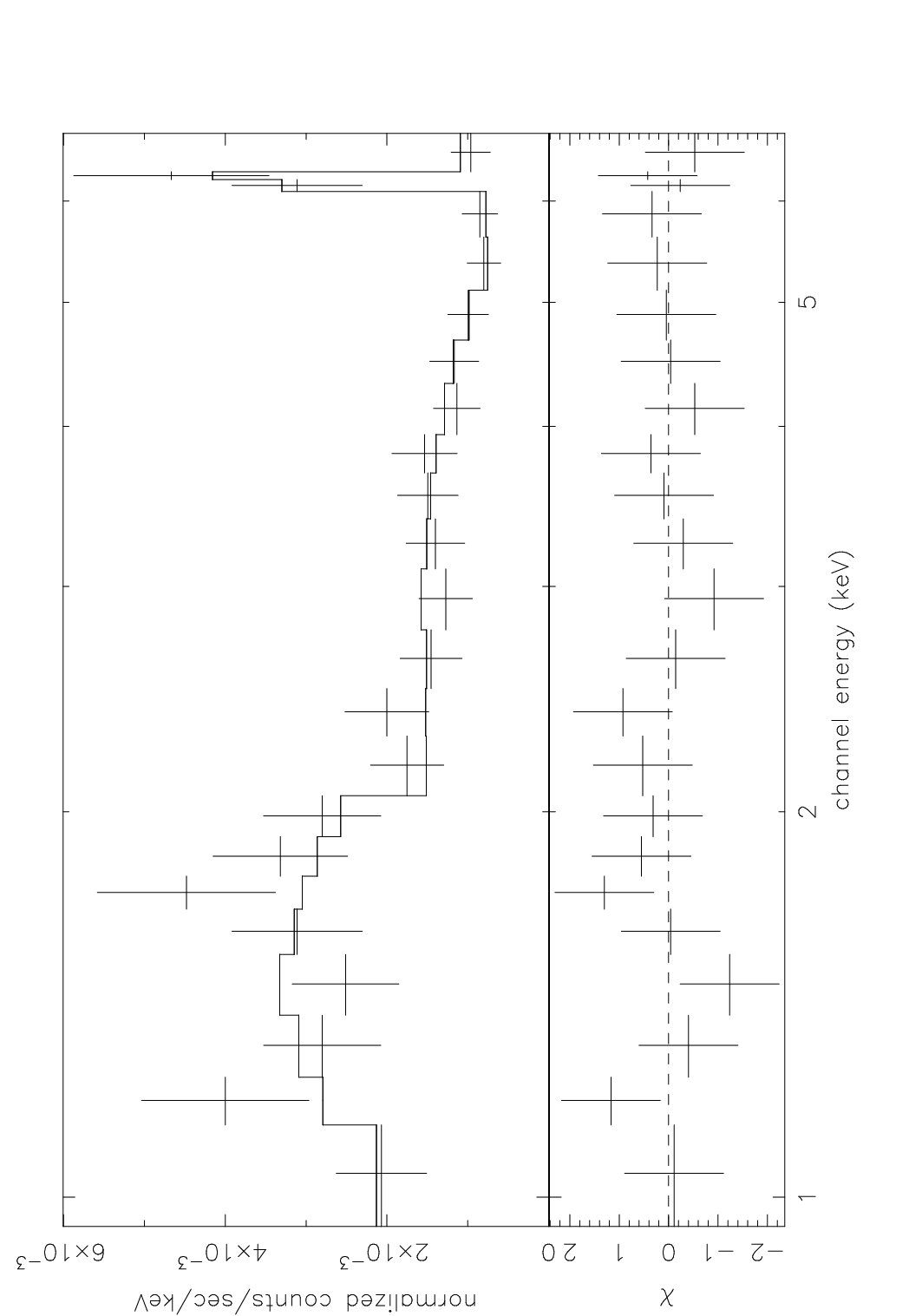}
\caption{Same as Fig. 4a, for northern nucleus.} 
\end{figure}

\end{document}